%% file: jakoubek.tex
\documentclass[11pt]{article}
\usepackage{xspace}
\usepackage{graphicx}
\usepackage{amsmath}
\usepackage{amssymb}
\usepackage{color}
\usepackage{subcaption}

\input econfmacros.tex
\def\Title#1{\begin{center} {\Large {\bf #1} } \end{center}}

\begin{document}

\Title{Flavour tagging techniques for CPV studies in the $B_s$ system with ATLAS}

\bigskip\bigskip


\begin{raggedright}  

Tomas Jakoubek\index{Jakoubek, T.}, {\it Institute of Physics ASCR, Prague, Czech Republic}\\

\begin{center}\emph{On the behalf of the ATLAS Collaboration.}\end{center}
\bigskip
\end{raggedright}

{\small
\begin{flushleft}
\emph{To appear in the proceedings of the 50 years of CP violation conference, 10 -- 11 July, 2014, held at Queen Mary University of London, UK.}
\end{flushleft}
}

\begin{abstract}
The sensitivity to $B_s$ meson mixing phenomena can be significantly improved by distinguishing the initial meson flavour. The flavour tagging techniques used in the ATLAS measurement of the angular amplitudes contributing to $B_s \rightarrow J/\psi\phi$ decays are presented using 4.9~fb$^{-1}$ of $pp$ data at $\sqrt{s} = 7$~TeV from 2011. The methods, their performance and calibration on self-tagged $B^\pm \rightarrow J/\psi K^\pm$ data are shown.
\end{abstract}

\section{Introduction}
In the Standard Model, $CP$ violation is described by a complex phase in the CKM matrix. One of the manifestations of this complex phase is a phase shift between direct and mixing-mediated $B_s$ decays producing a common final state. In the case of $B_s \rightarrow J/\psi\phi$ decay, this phase shift is predicted to be small: $\phi_s = 0.0368 \pm 0.0018$~rad \cite{Bona:2006sa}. New physics can enhance $\phi_s$ whilst satisfying all existing constraints. The sensitivity to $B_s$ meson mixing phenomena can be significantly improved by distinguishing the initial meson flavour.

In this paper, flavour tagging techniques used in the update \cite{Aad:2014cqa} to the previous untagged measurement \cite{Aad:2012kba} are presented. The analysis uses 4.9~fb$^{-1}$ of the LHC $pp$ data at $\sqrt{s} = 7$~TeV collected by the ATLAS detector in the year 2011. The weak phase $\phi_s$ and other parameters are extracted from the data by a simultaneous unbinned maximum likelihood fit to the $B_s \rightarrow J/\psi\phi$ candidates mass and time-dependent angular distributions.

\section{The ATLAS detector}
The ATLAS experiment \cite{Aad:2008zzm} is a multipurpose particle physics detector at the LHC, optimized for high $p_\mathrm{T}$ physics and instantaneous luminosity up to 10$^{34}$~cm$^{-2}$s$^{-1}$. The detector has a forward-backward symmetrical cylindrical geometry with an almost 4$\pi$ coverage. Precise tracking is provided by its innermost part, the Inner Detector (ID), which consists of a silicon pixel detector, a silicon microstrip detector and a transition radiation tracker, all immersed in a 2~T axial magnetic field. The ID covers the pseudorapidity region of $|\eta|<2.5$ and is enclosed by the electromagnetic and hadronic calorimeters. The outermost part of the ATLAS detector is the Muon Spectrometer (MS). It covers the pseudorapidity in the interval of $|\eta|<2.7$ and is located within the magnetic field produced by three large superconducting air-core toroid systems.

\section{Flavour Tagging}
Initial flavour of neutral $B$-mesons can be determined using information from the other $B$-meson typically produced from the other $b$-quark in the event. This technique is called the Opposite-Side Tagging (OST). From a calibration sample, the opposite-side charge is mapped to a probability $P$ that the event is a $B$ or $\bar{B}$. This probability is then put into the likelihood fit on a per-candidate basis. If no tagging information can be provided, $P = 0.5$ is assigned.

\subsection{Calibration Sample}
Self-tagging decay channel $B^\pm \rightarrow J/\psi K^\pm$ can be used to study and calibrate the OST methods. In this decay, flavour of the $B$-meson at production is provided by the kaon charge. Events from the entire 2011 run are used (integrated luminosity 4.9~fb$^{-1}$). $B^\pm$ candidates selection criteria:
\begin{itemize}
	\item the candidate is created from two oppositely-charged muons forming a good vertex
	\item each muon is required to have $p_\mathrm{T} > 4$~GeV and $|\eta| < 2.5$
	\item di-muon invariant mass must be in the range of $2.8 < m(\mu^+\mu^-) < 3.4$~GeV
	\item an additional track with the charged kaon mass hypothesis satisfying $p_\mathrm{T} > 1$~GeV, $|\eta| < 2.5$, is combined with the di-muon to form the $B$-candidate
	\item the $B$-candidate transverse decay length\footnote{Transverse decay length is the displacement in the transverse plane of the $B_s$ meson decay vertex with respect to the primary vertex, projected onto the direction of the $B_s$ transverse momentum.} must be $L_\mathrm{xy} > 0.1$~mm to reduce the prompt component of the combinatorial background
\end{itemize}

\begin{figure}[ht]
	\centering
	\includegraphics[width=0.6\columnwidth]{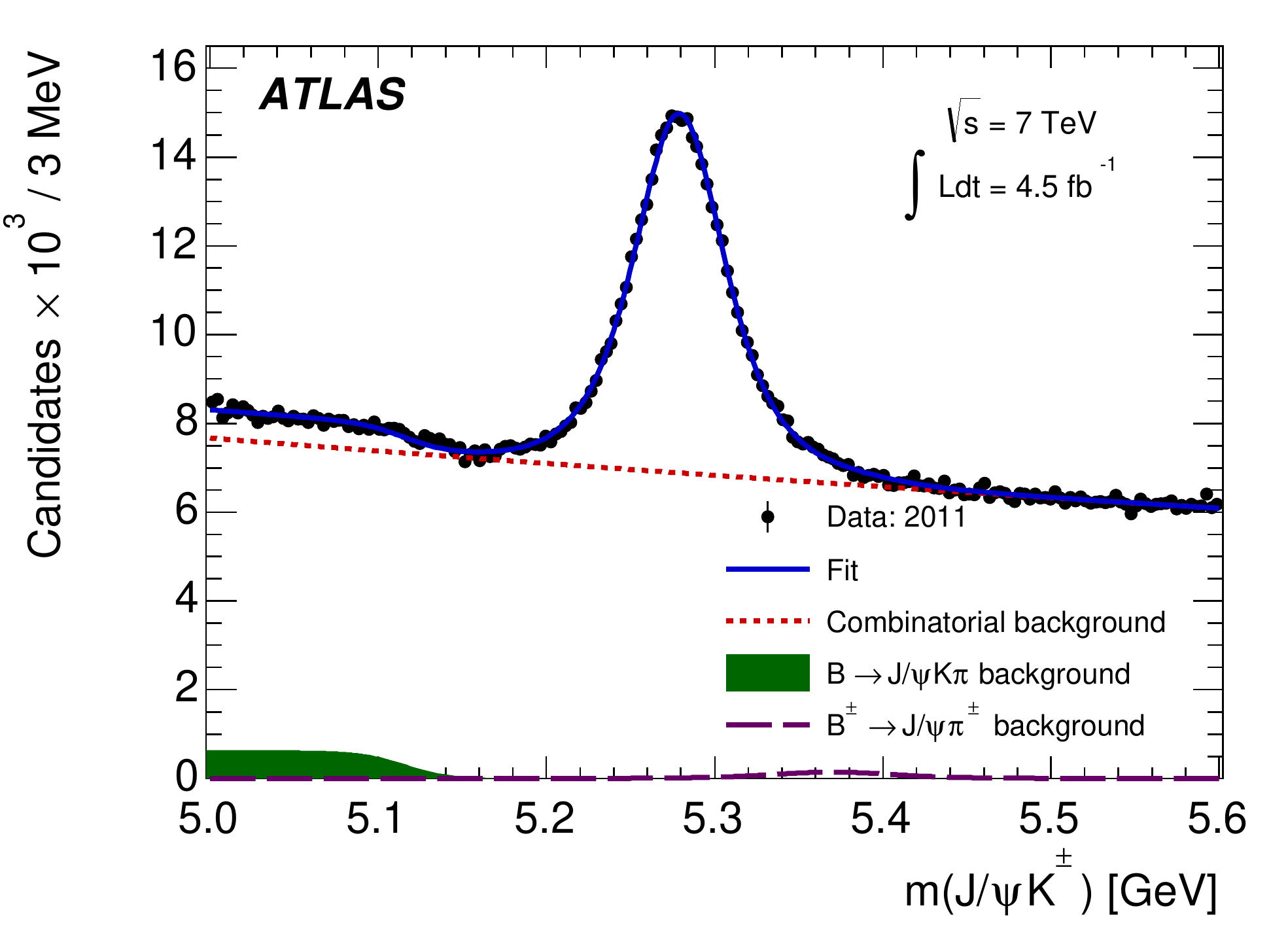}
	\caption{The invariant mass distribution for $B^\pm \rightarrow J/\psi K^\pm$ candidates. The blue curve represents the overall result of the fit, the red dotted line shows the combinatorial background component, partially reconstructed $B$ decays are shown as the green shaded area, and decays of $B^\pm \rightarrow J/\psi \pi^\pm$ (where the pion is mis-assigned a kaon) are given by the purple dashed line. The data are shown by black points. Taken from \cite{Aad:2014cqa}.}
	\label{fig:mass_Bpm}
\end{figure}

A sideband subtraction method is used to study the distributions of the signal processes with the background component removed. For this purpose, three mass regions are defined:
\begin{itemize}
	\item Signal region: around the fitted signal mass peak position $\mu \pm 2\sigma$
	\item Sidebands: $[\mu - 5\sigma, \mu - 3\sigma ]$ and $[\mu + 3\sigma, \mu + 5\sigma ]$
\end{itemize}
where $\mu$ and $\sigma$ are the mean and width of the Gaussian function describing the signal mass. Background is modelled by an exponential (combinatorial background), a hyperbolic tangent function (low-mass contribution from mis- and partially-reconstructed $B$ decays), and Gaussian function ($B^\pm \rightarrow J/\psi\pi^\pm$ contribution). Individual binned extended maximum likelihood fits are performed to the invariant mass distribution in five regions of $B$-candidate rapidity. The invariant mass distribution of $B^\pm$ candidates for all rapidity regions is shown in Figure~\ref{fig:mass_Bpm}.

\subsection{Flavour tagging using Muons}
\begin{figure}[ht]
        \centering
        \begin{subfigure}[b]{0.45\columnwidth}
                \includegraphics[width=\textwidth]{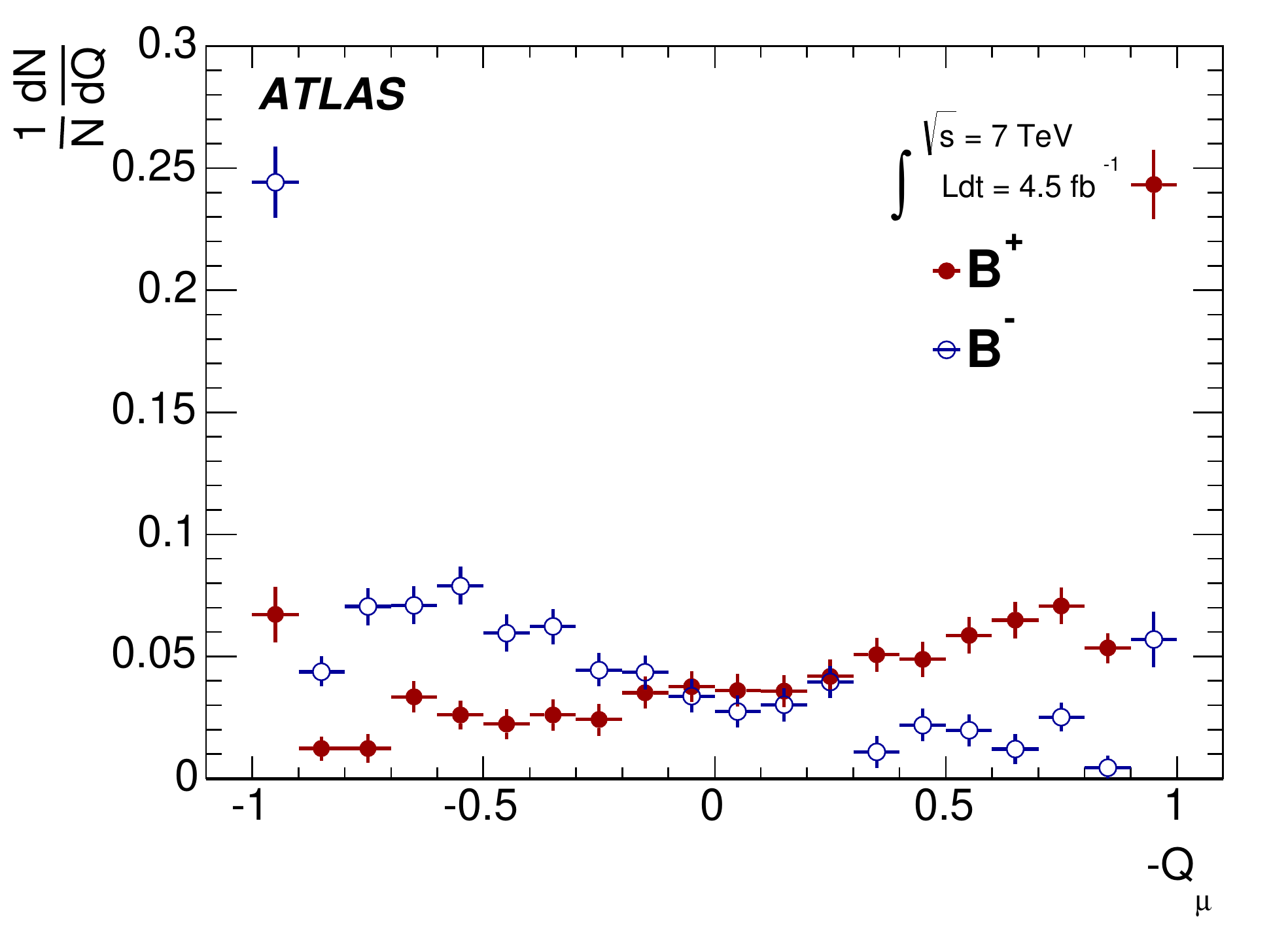}
        \end{subfigure}
        \begin{subfigure}[b]{0.45\columnwidth}
                \includegraphics[width=\textwidth]{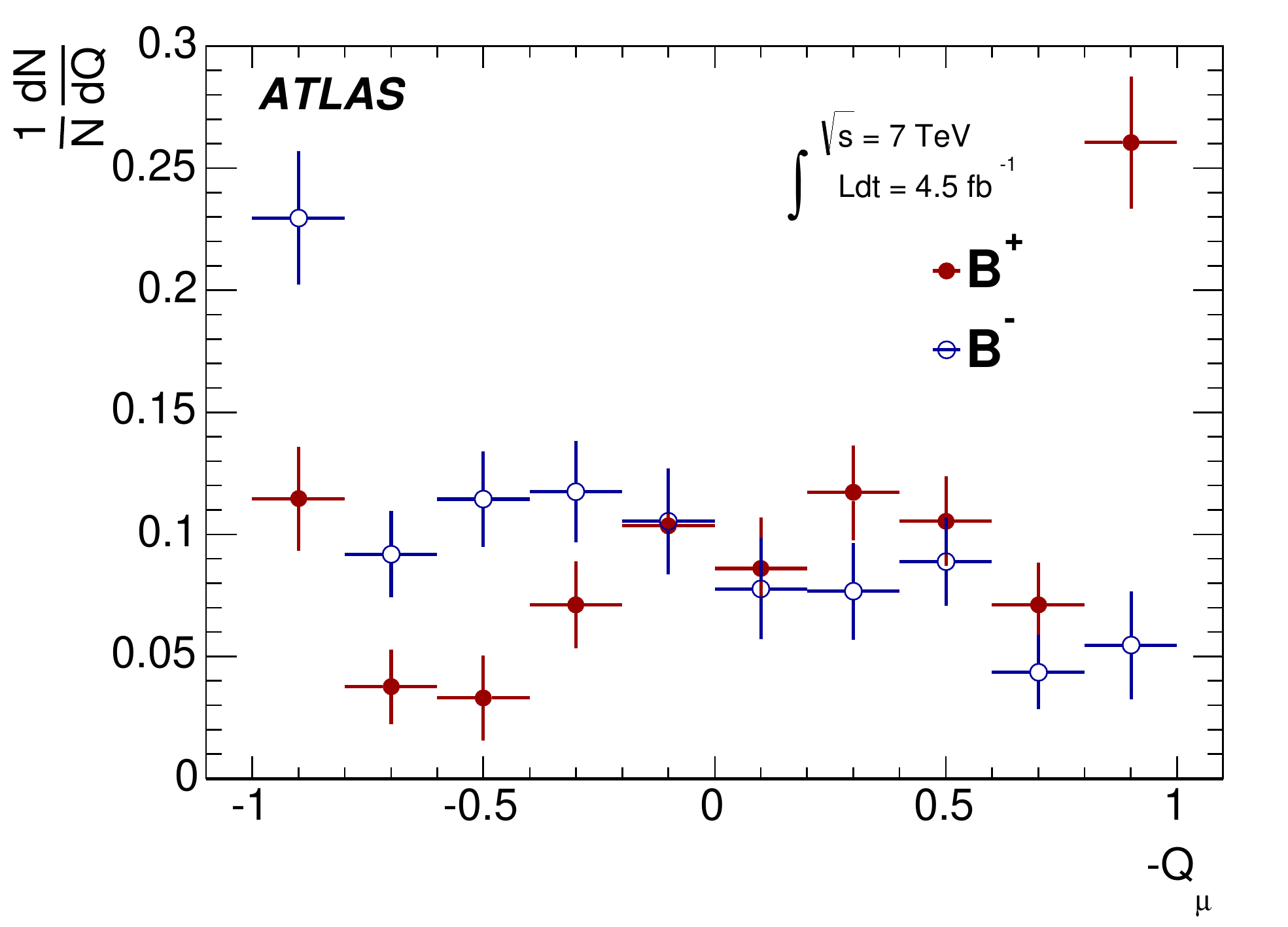}
        \end{subfigure}
        \caption{The opposite-side muon cone charge distribution from $B^\pm$ sample for combined muons (left) and segment tagged muons (right). Taken from \cite{Aad:2014cqa}.}
        \label{fig:tag_mu}
\end{figure}

One method to infer the flavour of the opposite-side $b$-quark is measuring the charge of a muon from the semileptonic decay of the $B$-meson. An additional muon is searched for in the event, having originated near the original interaction point. In the case of multiple muons, the one with the highest $p_\mathrm{T}$ is selected. Muons are separated into their two reconstruction classes\footnote{\textit{Combined muons} have a full track in the MS that is matched to a full track in the ID. \textit{Segment tagged muons} have a full track in the ID that is matched to track segments in the MS.}: \textit{combined} and \textit{segment tagged}. To enhance the separation power, muon \textit{cone charge} is defined
\begin{equation}
	Q_\mu = \frac{ \sum^{N\;\mathrm{tracks}}_i q^i \cdot (p_\mathrm{T}^i)^\kappa} { \sum^{N\;\mathrm{tracks}}_i(p_\mathrm{T}^i)^\kappa},
\end{equation}
where $\kappa = 1.1$ was tuned to optimise the tagging power; the sum is over the reconstructed tracks within a cone of $\Delta R < 0.5$ around the muon momentum, with $p_\mathrm{T} > 0.5$~GeV and $|\eta| < 2.5$ (tracks associated to the signal-side are excluded). The muon cone charge distributions from $B^\pm$ sample for both muon classes are shown in Figure~\ref{fig:tag_mu}.

\subsection{Flavour tagging using Jets}
In the absence of a muon, a b-tagged jet is required in the event, with tracks associated to the same primary vertex as the signal decay (excluding those from the signal candidate). The jet is reconstucted using the Anti-k t algorithm with a cone size of 0.6.

\begin{figure}[ht]
	\centering
	\includegraphics[width=0.45\columnwidth]{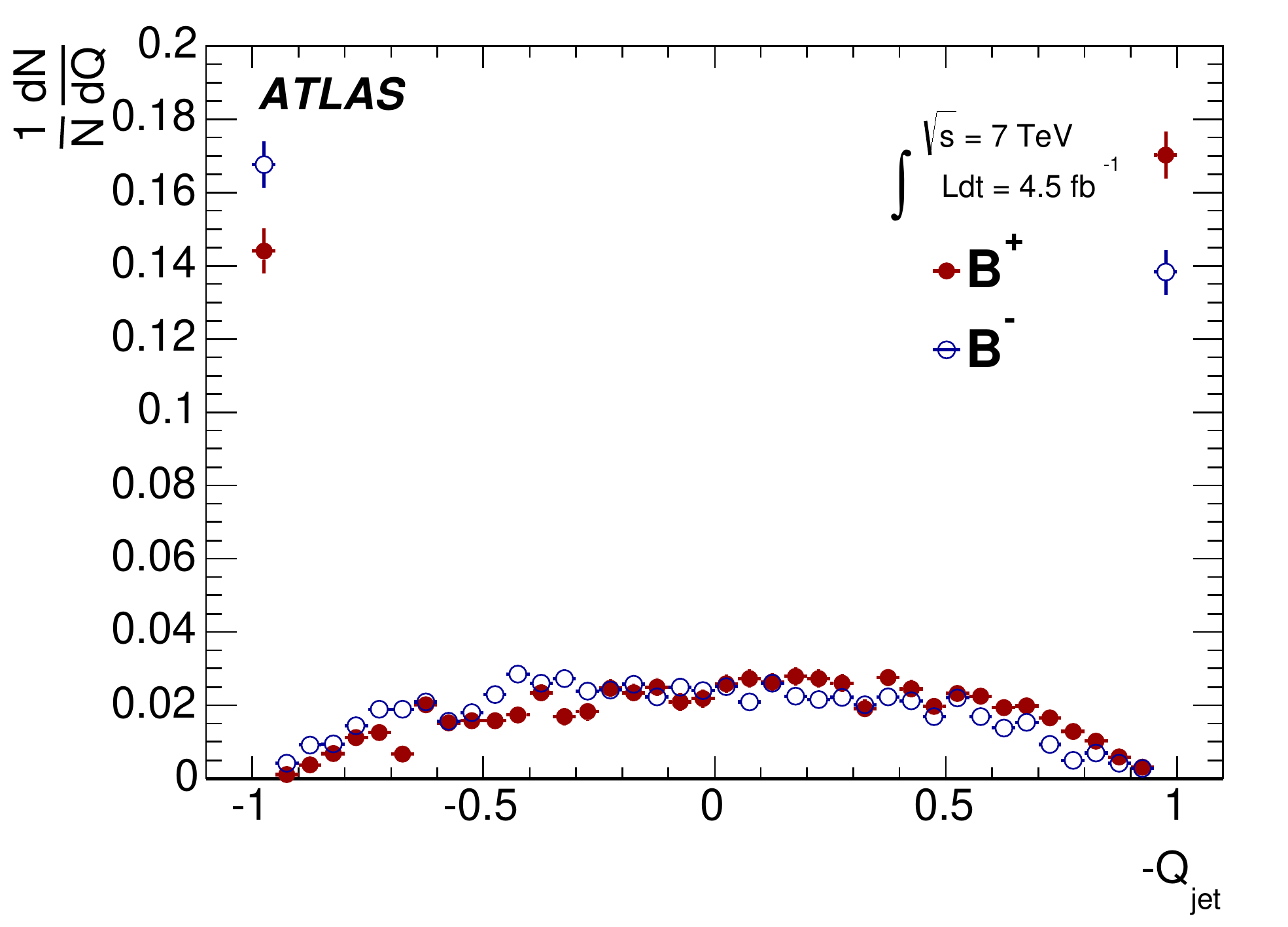}
	\caption{The opposite-side jet charge distribution from $B^\pm$ sample. Taken from \cite{Aad:2014cqa}.}
	\label{fig:tag_jet}
\end{figure}

Similarly to the muon cone charge, a \textit{jet charge} is defined
\begin{equation}
	Q_\mathrm{jet} = \frac{ \sum^{N\;\mathrm{tracks}}_i q^i \cdot (p_\mathrm{T}^i)^\kappa} { \sum^{N\;\mathrm{tracks}}_i(p_\mathrm{T}^i)^\kappa},
\end{equation}
where $\kappa = 1.1$, and the sum is over the tracks associated to the jet. The jet charge distribution from $B^\pm$ sample is shown in Figure~\ref{fig:tag_jet}.

\subsection{Tagging Probabilities}
Flavour tagging enters the fit in the form of the tag probability: a probability that a specific event has a signal decay containing a $\bar{b}$-quark, given the value of the discriminating variable  $P(B|Q)$ is constructed from the calibration samples for each of the $B^+$ and $B^-$ samples, defining $P(Q|B^+)$ and $P(Q|B^-)$ respectively. The probability to tag a signal event as containing a $\bar{b}$-quark is therefore $P(B|Q) = P(Q|B^+) / (P(Q|B^+) + P(Q|B^-))$ and $P(\bar{B}|Q) = 1-P(B|Q)$.

\begin{figure}[ht]
        \centering
        \begin{subfigure}[b]{0.45\columnwidth}
                \includegraphics[width=\textwidth]{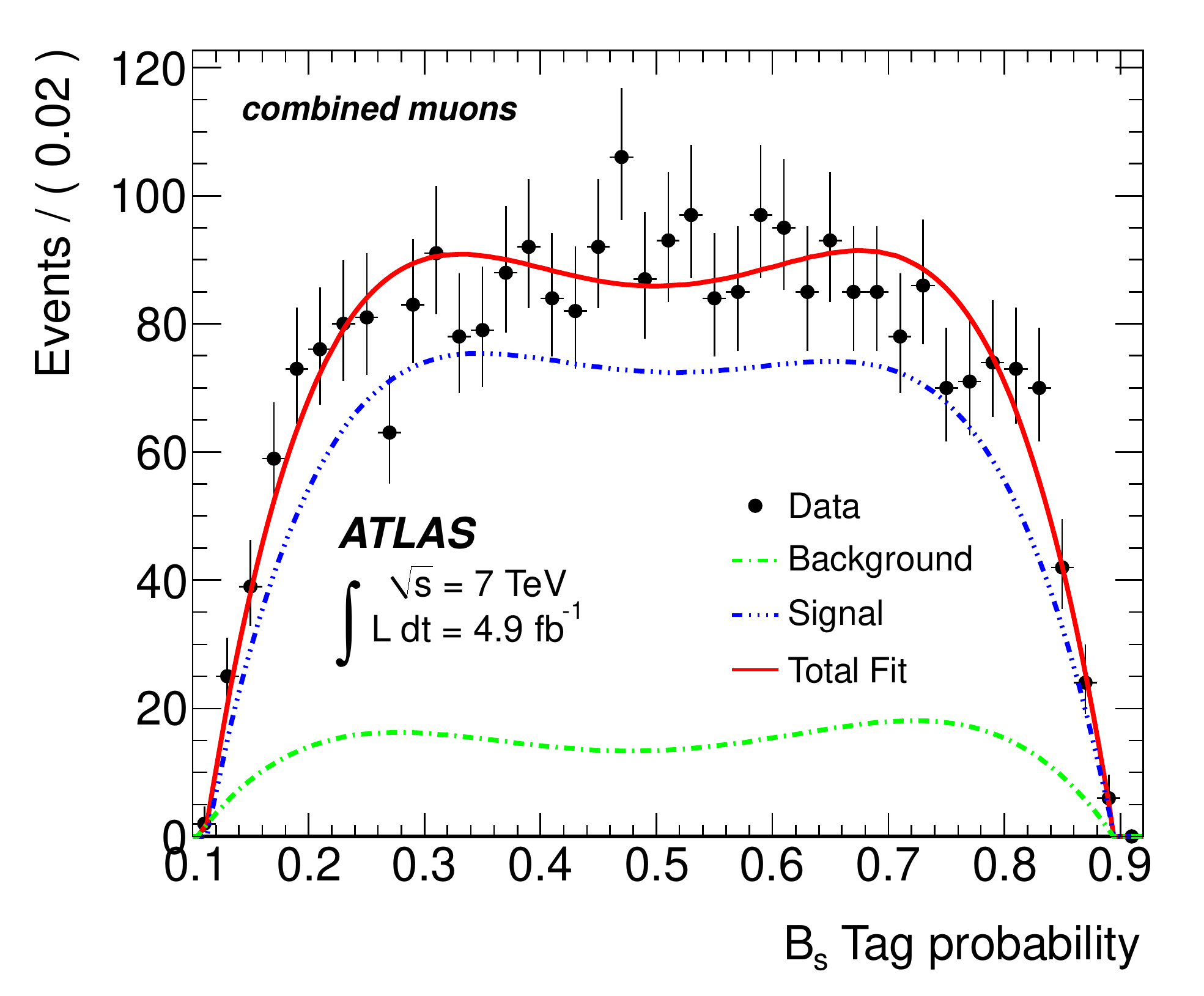}
                \caption{}
                \label{fig:tagprob_mu_comb}
        \end{subfigure}
        \begin{subfigure}[b]{0.45\columnwidth}
                \includegraphics[width=\textwidth]{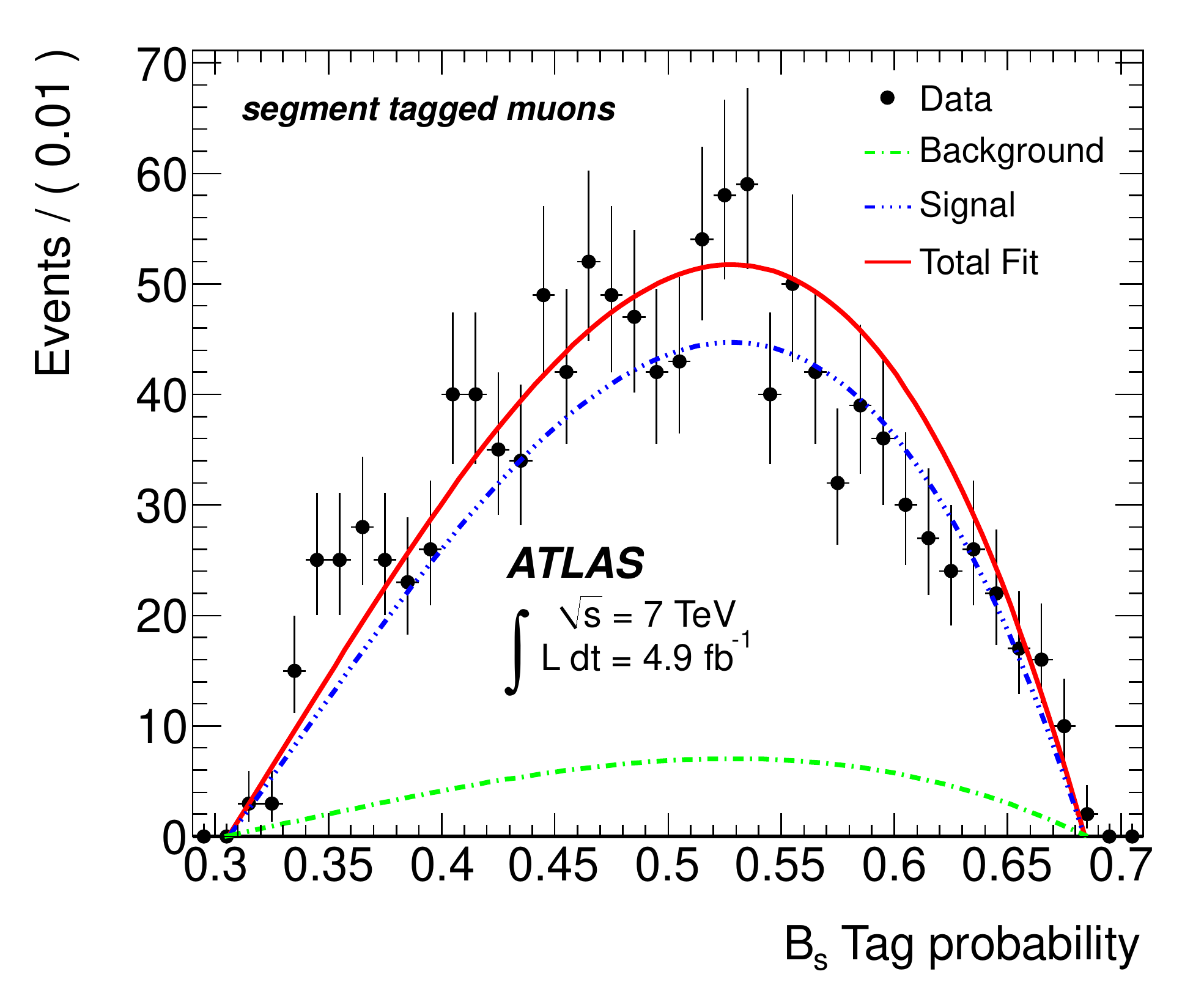}
                \caption{}
                \label{fig:tagprob_mu_seg}
        \end{subfigure}
        \begin{subfigure}[b]{0.45\columnwidth}
                \includegraphics[width=\textwidth]{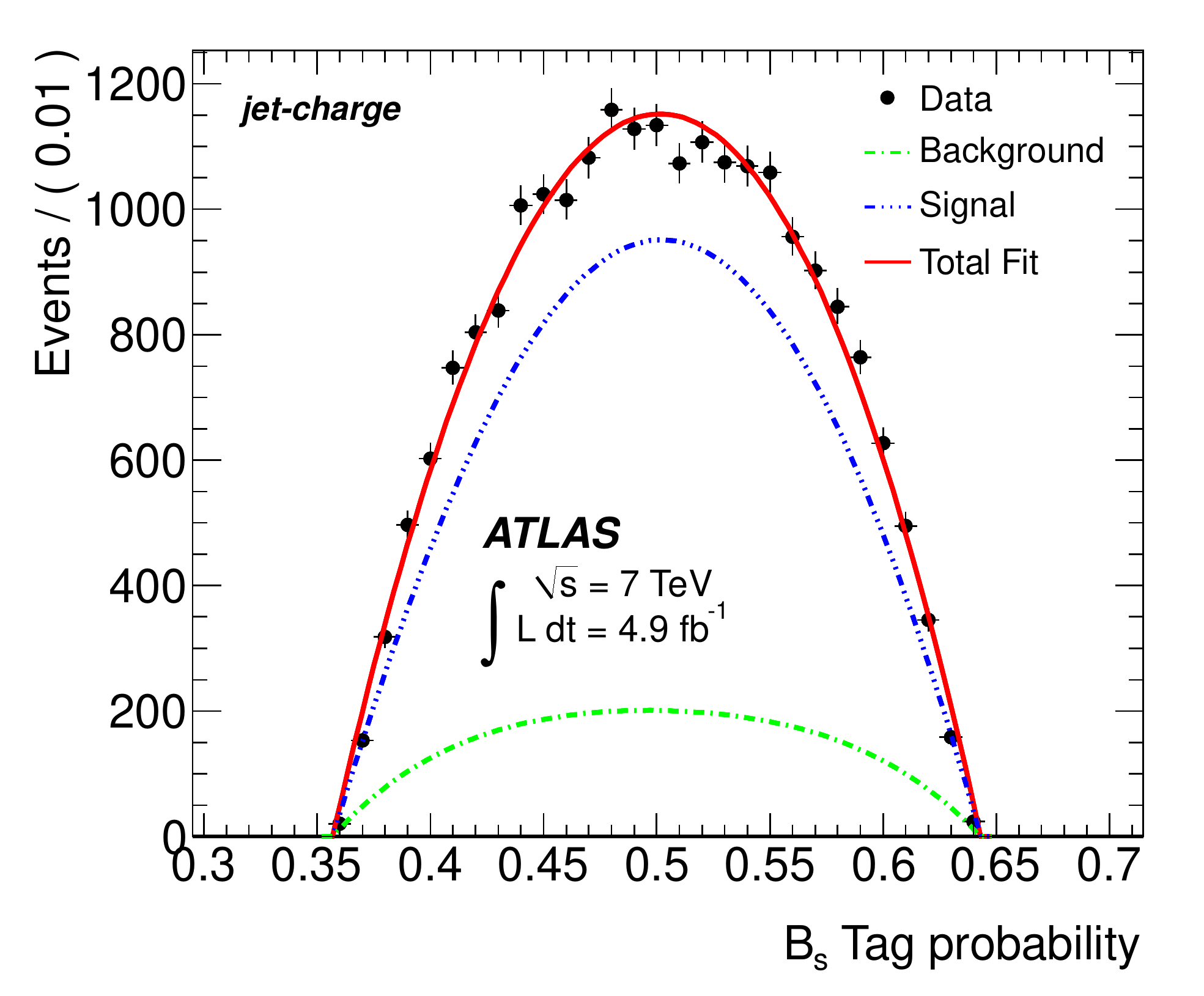}
                \caption{}
                \label{fig:tagprob_jets}
        \end{subfigure}
        \caption{Distributions of the $B_s$ tag probability obtained using combined muons (\subref{fig:tagprob_mu_comb}), segment tagged muons (\subref{fig:tagprob_mu_seg}), and jet charge (\subref{fig:tagprob_jets}). Black dots are data without the discrete parts, blue is the fit to the sidebands, green to the signal, and red is a sum of both fits. Taken from \cite{Aad:2014cqa}.}
        \label{fig:tagprob}
\end{figure}

Distributions of tag probabilities consist of continuous and discrete parts and are different for the signal and background. Since the background cannot be factorized out, extra PDF terms are included in the likelihood fit. These terms are obtained using a sidebands subtraction technique and fitting the continuous parts with (Chebychev) polynomial as shown in Figure~\ref{fig:tagprob}.

Discrete parts (not shown here) have their origin in tagging objects formed from a single track, providing a tag charge of exactly +1 or -1. For the fit, it is important to derive their fractions $f_{+1}$, $f_{-1}$. These can be found in Table~\ref{tab:tagprob}.

\begin{table}[ht]
\begin{center}
\caption{Summary of obtained relative probabilities between tag charges $+1$ and $-1$ for signal and background events for the different tagging methods. Only statistical errors are quoted.}
\begin{tabular}{l|cc|cc}
	\hline \hline
	Tag method           & \multicolumn{2}{c|}{Signal}           & \multicolumn{2}{c}{Background} \\
	                     & $f_{+1}$          & $f_{-1}$          & $f_{+1}$          & $f_{-1}$ \\
	\hline
	Combined $\mu$       & $0.106 \pm 0.019$ & $0.187 \pm 0.022$ & $0.098 \pm 0.006$ & $0.108 \pm 0.006$ \\
	Segment Tagged $\mu$ & $0.152 \pm 0.043$ & $0.153 \pm 0.043$ & $0.098 \pm 0.009$ & $0.095 \pm 0.008$ \\
	Jet charge           & $0.167 \pm 0.010$ & $0.164 \pm 0.010$ & $0.176 \pm 0.003$ & $0.180 \pm 0.003$ \\
	\hline \hline
\end{tabular}
\label{tab:tagprob}
\end{center}
\end{table}

\subsection{Tagging Performance}
The efficiency $\epsilon$ of an individual tagger is defined as the ratio of the number of tagged events to the total number of candidates. The tagging power is defined as
\begin{equation}
	{D}^2 = \sum_i \epsilon_i \cdot (2 P_i(B|Q_i) - 1)^2,
\end{equation}
where the sum is over the bins of the probability distribution as a function of the charge variable and $\epsilon_i$ is the number of tagged events in each bin divided by the total number of candidates. An effective dilution $\mathcal{D}$ is calculated from the tagging power and efficiency.

The combination of tagging methods is applied according to the hierarchy of performance, based on the dilution of the tagging method (see Table~\ref{tab:tagperf}).
\begin{table}[ht]
\begin{center}
\caption{Summary of tagging performance for the different tagging methods. Only statistical errors are quoted.}
\begin{tabular}{l|c|c|c}
	\hline \hline
	Tagger               & Efficiency $[\%]$ & Dilution $[\%]$  & Tagging Power $[\%]$ \\
	\hline
	Combined $\mu$       & $3.37 \pm 0.04$   & $50.6 \pm 0.5$   & $0.86 \pm 0.04$ \\
	Segment Tagged $\mu$ & $1.08 \pm 0.02$   & $36.7 \pm 0.7$   & $0.15 \pm 0.02$ \\
	Jet charge           & $27.7 \pm 0.1$    & $12.68 \pm 0.06$ & $0.45 \pm 0.03$ \\
	\hline
	Total                & $32.1 \pm 0.1$    & $21.3 \pm 0.08$  & $1.45 \pm 0.05$ \\
	\hline \hline
\end{tabular}
\label{tab:tagperf}
\end{center}
\end{table}

\section{Summary}
\begin{figure}[ht]
        \centering
        \begin{subfigure}[b]{0.45\columnwidth}
                \includegraphics[width=\textwidth]{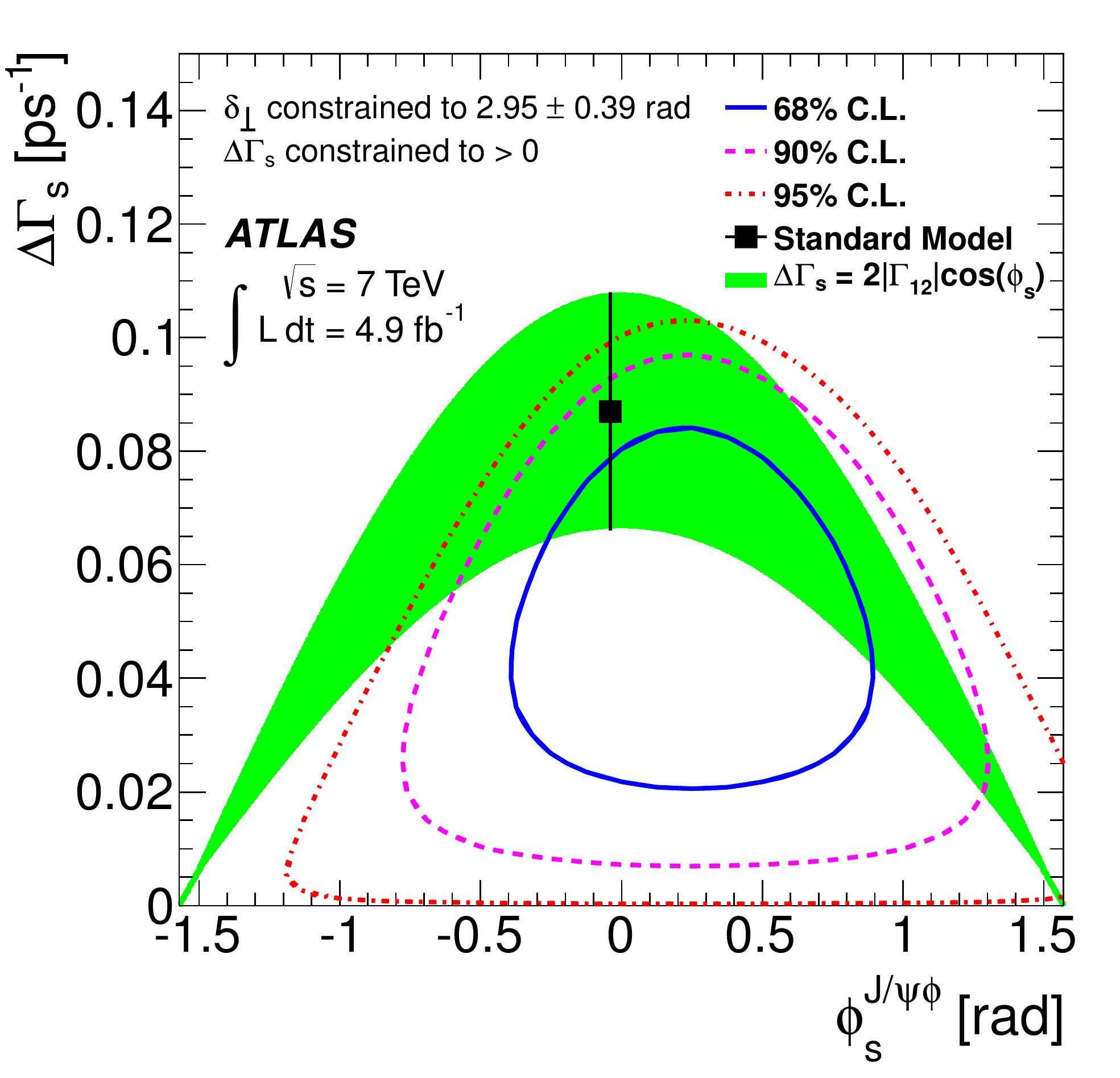}
                \caption{}
                \label{fig:cont_untagged}
        \end{subfigure}%
        \begin{subfigure}[b]{0.45\columnwidth}
                \includegraphics[width=\textwidth]{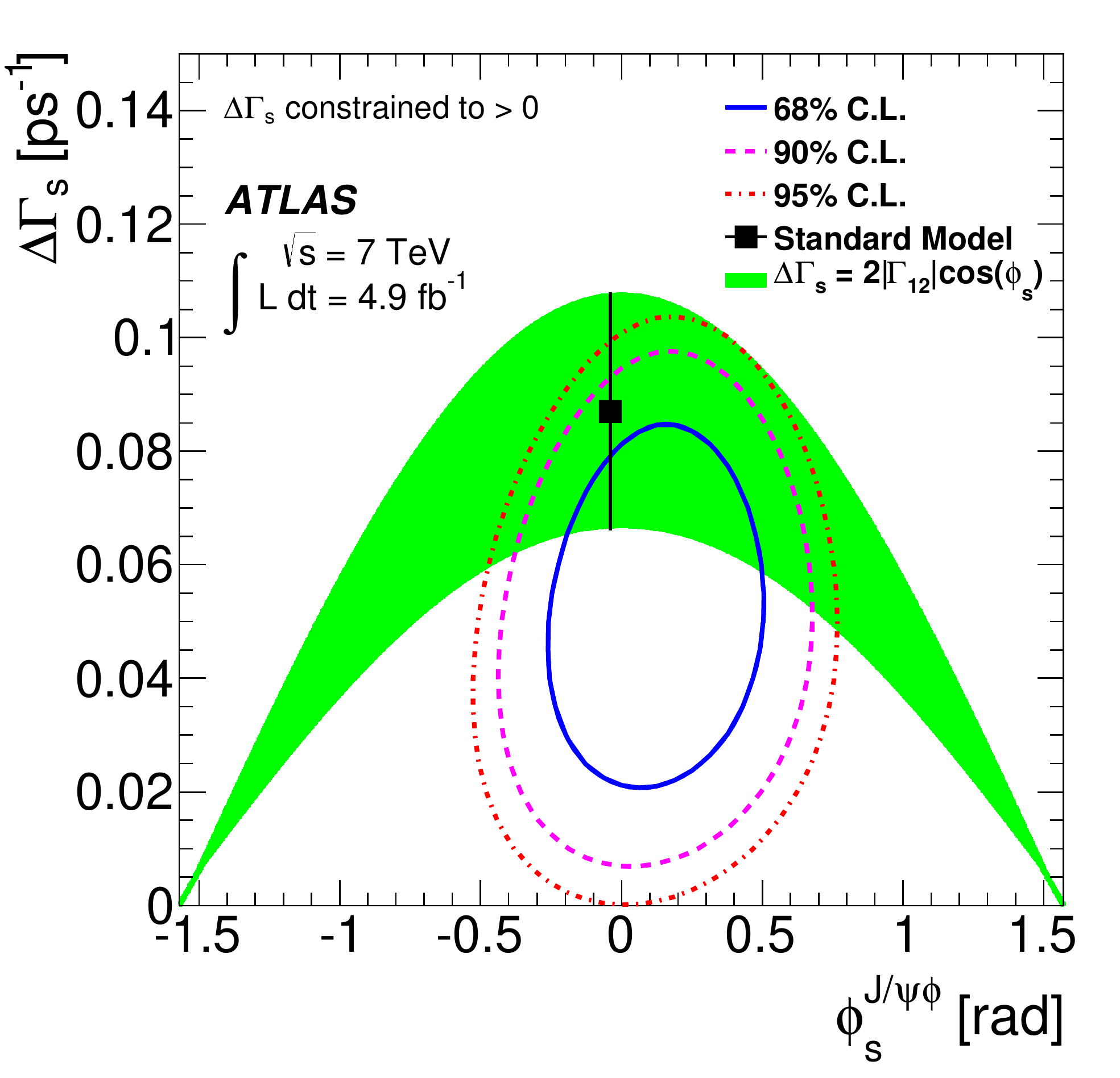}
                \caption{}
                \label{fig:cont_tagged}
        \end{subfigure}
        \caption{Likelihood contours in the $\phi_s$ -- $\Delta\Gamma_s$ plane for (\subref{fig:cont_untagged}) untagged (taken from \cite{Aad:2012kba}) and (\subref{fig:cont_tagged}) tagged (taken from \cite{Aad:2014cqa}) 2011 data analysis. The blue line shows the 68\% likelihood contour, the dashed pink line shows the 90\% likelihood contour and the red dotted line shows the 95\% likelihood contour (statistical errors only). The green band is the theoretical prediction of mixing-induced $CP$ violation.}
        \label{fig:cont}
\end{figure}

An update to our untagged analysis of $B_s \rightarrow J/\psi\phi$ decay using 4.9~fb$^{-1}$ of ATLAS 2011 data was performed. Using the flavour tagging techniques has significantly improved the overall uncertainty of the weak phase $\phi_s$:
\begin{eqnarray*}
	\mathrm{Untagged:} & \phi_s = 0.22 \pm 0.41\ (\mathrm{stat.}) \pm 0.10\ (\mathrm{syst.})\ \mathrm{rad} \\
	\mathrm{Tagged:} & \phi_s = 0.12 \pm 0.25\ (\mathrm{stat.}) \pm 0.05\ (\mathrm{syst.})\ \mathrm{rad}.
\end{eqnarray*}
Other measured values are consistent with those obtained in our previous analysis and also with the values predicted in the Standard Model. A comparison of the contour plots in the $\phi_s$ -- $\Delta\Gamma_s$ plane for untagged and tagged analysis is shown in Figure~\ref{fig:cont}.

\bigskip
\section{Acknowledgments}
I would like to thank for support from grants of the Ministry of Education, Youth and Sports of the Czech Republic under the project LG13009.

\end{document}

%% file: econfmacros.tex
\definecolor{Red}{rgb}{1,0,0}
\definecolor{Green}{rgb}{0,1,0}
\definecolor{Blue}{rgb}{0,0,1}
\definecolor{Black}{rgb}{0,0,0}

\def\beq{\begin{equation}}
\def\eeq#1{\label{#1}\end{equation}}
\def\eeqn{\end{equation}}

\def\beqa{\begin{eqnarray}}
\def\eeqa#1{\label{#1}\end{eqnarray}}
\def\eeqan{\end{eqnarray}}

\let\bar=\overbar

\def\Dslash{\not{\hbox{\kern-4pt $D$}}}
\def\dslash{\not{\hbox{\kern-2pt $\del$}}}

\def\msb{{\bar{\ssstyle M \kern -1pt S}}}